\documentstyle[12pt,aasms4,epsf,rotate]{article}

\def\kpc{ {\rm kpc} }

\def\pc{ {\rm pc} }

\def\imin{I_{\rm min}}
\def\imax{I_{\rm max}}
\def\msun{{\rm M}_{\rm \odot}}
\def\Rsun{R_{\rm \odot}}

\def\sech{{\rm sech}}
\def\zthin{z_{\rm thin}}
\def\zthick{z_{\rm thick}}
\def\zthinz{{\zthin}_{,\odot}}
\def\zthickz{{\zthick}_{,\odot}}

\def\spose#1{\hbox to 0pt{#1\hss}}
\def\lta{\mathrel{\spose{\lower 3pt\hbox{$\sim$}}
    \raise 2.0pt\hbox{$<$}}}
\def\gta{\mathrel{\spose{\lower 3pt\hbox{$\sim$}}
    \raise 2.0pt\hbox{$>$}}}

\begin{document}

\title{Star Counts and the Warped and Flaring Milky Way Disk}
\author{Geza Gyuk}
\affil{S.I.S.S.A., via Beirut 2-4, 34014 Trieste, Italy\altaffilmark{1}\altaffiltext{1}{
	present address: Department of Physics, UCSD, 9500 Gilman Dr.,
La Jolla, CA 92093}}
\and
\author{Chris Flynn}
\affil{Tuorla Observatory, Piikkio, FIN 21500, Finland}
\and
\author{N. Wyn Evans}
\affil{Theoretical Physics, Department of Physics, 1 Keble Rd, Oxford,
OX1 3NP, UK}

\begin{abstract} 
This paper reports the star count predictions of warped and flaring
models of the outer Milky Way disk. These have been suggested as
possible locations of the lenses responsible for the microlensing
events towards the Large Magellanic Cloud (LMC). Three such models are
investigated in detail and the theoretical predictions are confronted
with {\it Hubble Space Telescope} (HST) star count data in 7 low
latitude fields ($30^\circ < | b| < 40^\circ$).  If the warped and
flaring disk population has the same characteristics as disk stars in
the solar neighbourhood, then the models can be unambiguously ruled
out. Metallicity gradients are well-known in disk galaxies and may
cause the outer disk population to differ in colors and luminosity
from that locally. This effect is studied using a simple ansatz for
the variation in the color-magnitude relation with position and while
it is shown to lead to better agreement with the star counts, upper
limits on the contribution of the warped and flaring disk to the
optical depth are still below the value measured towards the LMC.
Only if the warp is very asymmetric or if the luminosity function
changes strongly with Galactocentric radius can the models be made
consistent with the starcounts.
\end{abstract}


\section{INTRODUCTION}

The difficulty of explaining the origin of the microlensing events
towards the Large Magellanic Cloud (LMC) with conventional models of
the Milky Way has led to the suggestion that the outer parts of the
stellar disk may be warped and flared (Evans, Gyuk, Turner \& Binney
1998). Direct observational evidence on the extent of warping and
flaring in the outer Milky Way is hard to gain, but models in which
the stars warp and flare as strongly as the neutral gas can provide
microlensing optical depths of $\sim 1 \times 10^{-7}$.  The aim of
this paper is to examine this suggestion with a critical eye in the
light of the new deep star count data that is available from the Wide
Field Camera on the Hubble Space Telescope (e.g., Gould, Bahcall \&
Flynn 1998).  The warp of the stellar disk is difficult to test
directly. This is because searching closer than about $20^\circ$ to
the Galactic plane is difficult as the confusion due to regular
foreground sources becomes too high. By comparison, the maximum warp
angle is only about $8^\circ$, and then only along the ridge-line of
the warp. An easier target is to search directly for the evidence of
the flare that is assumed on top of the warp. It is the flare that is
mainly responsible for increasing the microlensing optical depth
towards the LMC to values broadly consistent with the observations.
The flare must put an enhanced number of stars along the line of sight
to the LMC ($\ell = 280^\circ, b = -33^\circ$). In this paper, our aim
is to confront the theoretical predictions of warped and flaring disk
models with the star count data on low latitude fields with $30^\circ
< | b| < 40^\circ$.

Section 2 introduces the dataset under scrutiny. In Section 3, the
simplest assumption is made -- namely that the stellar population in
the warped and flaring disk has the same luminosity function as that
determined for the local disk. Section 4 considers more elaborate
models in which the effects of metallicity gradients are taken into
account, albeit crudely.

\section{THE STAR COUNT DATASET}

The dataset is part of an on-going star count project being conducted
with the Wide Field Camera on the HST (see e.g., Gould, Bahcall \&
Flynn 1996, 1997, 1998).  The seven fields we have selected for
scrutiny are listed in Table 1a.  They have been chosen so that
$30^\circ < | b| < 40^\circ$ and so are optimum for our purposes of
tracing the stellar flare at the lattitude of the Magellanic
Clouds. The table also lists the $I$ band saturation magnitude $\imin$
and the detection threshold $\imax$. The saturation magnitude varies
from field to field as it depends on the exposure time of the shortest
exposure used in producing the cosmic ray cleaned images. The
detection threshold also varies from field to field -- it is the
faintest magnitude at which stars and galaxies can still be
unambiguously distinguished. Full details of the methods of
determining these limits are given in Gould, Bahcall \& Flynn
(1996). To be conservative, 1.5 magnitudes are added to the nominal
bright limit and 0.3 magnitudes are subtracted from the nominal faint
limit. On the faint side, this guarantees absolute safety with regard
to star/galaxy separation. At the bright end, the 1.5 magnitudes are
due to technical reasons concerning the pixel size.  Only data points
between these new limits are actually used in constructing the star
count statistics.

Table 1b gives the star count data in the color range $1.52 < V-I <
2.26$ in the seven fields. This selection function picks out M dwarfs,
as discussed in Gould, Bahcall \& Flynn (1996). The M dwarfs should be
representative of the disk, and so should presumably participate in
any outer flaring or warping. Table 1b shows that each field contains
only a few stars. The numbers of stars in each of seven magnitude bins
is recorded. The bins are of 1 magnitude extent. The $I$ band
magnitude centroids of the bins are indicated at the top of each
column.

\section{WARPED AND FLARING DISK MODELS}

The density of the warped and flaring disk model as a function of
galactocentric radius $R$ and height $z$ above the Galactic plane is:
\begin{equation} 
\rho(R,z) = \rho_0 \exp (-R/ R_d) \Bigl[ (1-f) \exp (-|z - z_w|/
\zthick ) + f \sech^2 ( |z - z_w|/ \zthin ) \Bigr].  
\end{equation} 
The overall normalisation $\rho_0$ is chosen to reproduce the column
density of the thin and thick disks at the solar radius
($\Rsun=8\,$kpc) of $46 \pm 3\msun \pc^{-2}$, in agreement with recent
dynamical estimates (Gould 1990, Holmberg \& Flynn 1998).  More
important from the standpoint of comparison with star counts is the
normalisation of the disk with the contribution of the gas ($\sim 13
\msun \pc^{-2}$) and the disk white dwarfs ($\sim 4
\msun \pc^{-2}$) excluded. This column density is taken as 
$\sim 30 \msun \pc^{-2}$. (Let us remark that this value inferred from
dynamics is perhaps some $\sim 15 \%$ higher than suggested by Gould,
Bahcall \& Flynn (1997) on the basis of starcounts. These authors
reckon that the total column density in stars is greater than the
column density in M dwarfs by a factor of $2.1$, leading to a value of
$\sim 26 \msun \pc^{-2}$).

The vertical profile of the thin disk follows roughly a
hyperbolic-secant squared profile whereas the thick disk is modelled
by an exponential profile in height above or below the midplane of the
warp $|z-z_w|$. The fractional contributions of the thin and thick
disk to the midplane density are denoted by $f$ and $1-f$.  The sun
lies on the line of nodes of the warp, which is described by the
equation:
\begin{equation}
z_w (R, \theta) = 0.214 \times (R - \Rsun) \sin \theta
\end{equation}
and sets in for radii $R$ greater than the solar position $\Rsun$.
Here, $\theta$ is the longitude measured about the Galactic Center
from the line of nodes. Observationally, the stellar warp cannot be
detected to large radii. Our equation is suggested by the behaviour of
the gas warp with one important difference. The gas warp bends back to
the Galactic plane in the southern hemisphere for $R \gta 15\,\kpc$
(Burton 1988), but the amplitude of the stellar warp is assumed to
continue increasing linearly. The scaleheights of the thin and thick
disks flares by a factor $F$ between the solar position and $25\,\kpc$
and so are described by;
\begin{equation}
\zthin = \zthinz \Bigl[1+ (F-1){R - \Rsun\over 25 - \Rsun}\Bigr],
\qquad\qquad
\zthick = \zthickz \Bigl[ 1+ (F-1){R - \Rsun\over 25 - \Rsun}\Bigr].
\end{equation}
For comparison, the scaleheight of the neutral gas is known to flare
by a factor $F$ of roughly ten over this distance (e.g., Wouterloot,
Brand, Burton \& Kwee 1990). Table 2 gives the scale length, flaring
factor, relative normalization of the thin and thick disks and the
scale heights for the 3 models representative models we consider
here. For all models the disk is truncated at 25 kpc. Also shown is
the optical depth towards the LMC. Effects such as ellipticity of the
disk, or lopsidedness, which can increase the optical depth further
are not considered.

Given the density law, a luminosity function $\Phi$ is needed to make
star count predictions. The M dwarf luminosity function locally is
recorded in Fig. 2 of Gould, Bahcall \& Flynn (1998). Between $8 < M_v
< 10.5$, the $V$ band luminosity function can be approximated by the
linear fit
\begin{equation}
\Phi = 0.0633\times \Bigl[ 3.16(M_v - 8.) + 1.2 \Bigr],
\end{equation}
where $\Phi$ is normalised to unit density. As the star counts
are known in the $I$ band, but the luminosity function is readily
available in the $V$ band, we will need to convert using the
disk color-magnitude relation (Gould, Bahcall \& Flynn 1998)
\begin{equation}
M_v = 2.89 + 3.37(V-I)
\end{equation}
and assuming vanishing extinction so that $V -I = M_v - M_I$.  The
results for our three models are given in Table 3. It is clear that if
the characteristics of the outer Milky Way disk are the same as those
of the local solar neighborhood, all three models are strongly ruled
out. This conclusion has been confirmed by Holmberg (1998, private
communication) in a starcount analysis of a somewhat similar model --
namely the modified Bahcall-Soneira model discussed in Holmberg, Flynn
\& Lindegren (1997) on which a flare and warp have been superposed.

Neglecting the contribution of the spheroid, we perform a maximum
likelihood fit of the overall normalisation of the stellar disk to the
star count data. The fit is best -- though still poor -- for a disk
surface density (participating in the flare) of $\approx 2.5 \msun
\pc^{-2}$. This gives an upper limit to the optical depth available
from the warped and flaring disk of $\approx 3 \times 10^{-9}$.

\section{THE EFFECT OF COLOR GRADIENTS}

The assumption that the disk stellar population is constant with
radius is attractive from the point of view of simplicity.
Unfortunately, it is hardly likely to be correct.  Color, metallicity
and even mass function gradients are well established in disk galaxies
(Pagel 1998, Taylor 1998). Spheroid stars in this color range are
typically 2.5 magnitudes fainter than disk stars of the same color
(Monet et al. 1992).  All this suggests investigation of
color-magnitude relations whose properties change with Galactocentric
radius such as
\begin{equation}
M_v = 2.89 + 3.37(V-I) + 1.66 \lambda (R - \Rsun)
\end{equation}
Here, $\lambda$ is the metallicity gradient in dex kpc{}$^{-1}$.  We
have assumed that the change in magnitude between disk and spheroid is
$\Delta M_v = 2.5$ and the corresponding change in metallicity is
$\Delta Z = 1.5$. Following the findings of Hippel et al. (1996), we
assume that the luminosity function for the range we are interested in
is independent of metallicity.

The results of adding a metallicity gradient can be seen in Table 4
which shows the star counts expected for model A with $\lambda = 0.15$
dex kpc${}^{-1}$. Clearly, the number of stars expected has been
substantially reduced. However, performing the same maximum likelihood
fit as previously, we find that the upper limit on the optical depth
has only improved $\approx 10^{-8}$, still far below the measured
optical depth towards the LMC.  This is perhaps hardly surprising. It
is well-known that the deep star count data are consistent with a
spheroid whose density falls off like $r^{-3.5}$ and which contributes
at most a few percent to the optical depth. If a warped and flaring
disk is to be consistent with the same star count data, its optical
depth must be nearly the same to within a geometrical factor of a few.

\section{CONCLUSIONS}

The warped and flaring disk models of the outer parts of the Milky Way
cannot provide a substantial contribution to the microlensing optical
depth and still be consistent with the starcount data. This statement
needs two qualifications. First, the starcounts presented here are not
exactly in the direction of the Large Magellanic Cloud (LMC). Strictly
speaking, it is therefore the assumed azimuthal variation of the warp
and flare that is being directly tested. It remains possible that a
strongly asymmetric warp (perhaps generated directly by the LMC) can
be made consistent with the starcounts. Second, the luminosity
function or mass function may vary strongly with radius and this may
allow the lenses to remain dark enough to evade detection even in very
deep fields. This idea has sometimes been regarded as unappealing when
measured against the yardstick of theoretical simplicity, though it
has recently been resuscitated by Kerins \& Evans (1998) and Binney
(1998). In particular, there may be two modes of star formation
controlled by metallicity. It might be that at low metallicity, only
low mass objects are formed, whereas at higher metallicity, mainly
viable stars are produced.

Many researchers have suggested that the microlensing events seen
towards the LMC may occur not in the Galactic halo but in intervening
structures. Possibilities include not just the warped and flaring disk
of the Galaxy, but also stellar conglomerations in front of the
Magellanic Clouds (Zaritsky \& Lin 1997, Zhao 1998) and hitherto
undetected populations in the outer parts of the Galaxy (Gates, Gyuk,
Holder \& Turner 1998). If these populations are traced by at least
some bright stars, then they should be visible in deep starcounts
towards the LMC. So, all these suggestions are very vulnerable to
direct observational testing both with the {\it Hubble Space
Telescope} and the new generation of large telescopes (such as the
{\it Very Large Telescope}).

\acknowledgments GG thanks the Theoretical Physics Department for
hospitality during a working visit to Oxford University, the Department of
Energy for partial support under grant DEFG0390ER40546 and Research
Corporation.  NWE is supported by the Royal Society. All three authors owe
a debt of gratitude to John Bahcall and Andy Gould for permission to use
their star count data in advance of publication. We also thank Johan
Holmberg for telling us of his own investigations into the problem.

\eject

\begin{table*}
\begin{center}
\begin{tabular}{crrrrrrrrrr}
Field No & $\ell$ & $b$ & $\imin$ & $\imax$ \\
\tableline
4 & $179.828^\circ$ & $-32.146^\circ$ & $17.2$ & $23.9$ \\
36 & $56.716^\circ$ & $34.251^\circ$ & $18.3$ & $24.2$ \\
38 & $56.726^\circ$ & $34.248^\circ$ & $18.3$ & $24.2$ \\
71 & $149.784^\circ$ & $34.678^\circ$ & $18.0$ & $23.8$ \\
72 & $149.784^\circ$ & $34.712^\circ$ & $18.4$ & $24.0$ \\
94 & $303.348^\circ$ & $33.66^\circ$ & $18.3$ & $24.1$ \\
140 & $77.018^\circ$ & $35.78^\circ$ & $18.1$ & $24.2$ \\
\end{tabular}
\end{center}


\tablenum{1a}
\caption{This table lists the seven fields under scrutiny. The first
column gives the (arbitrary) field identifier and the next columns give
the location in terms of Galactic longitude and latitude. The final
two columns give the limiting $I$ band magnitudes.}

\end{table*}

\eject

\begin{table*}
\begin{center}
\begin{tabular}{crrrrrrrrrr}
Field No & $18$ & $19$ & $20$ & $21$ & $22$ & $23$ & $24$ \\
\tableline
4  &  $0$ & $0$ & $2$ & $0$ & $1$ & $0$ & $0$ \\
36 &  $0$ & $0$ & $1$ & $1$ & $3$ & $3$ & $1$ \\
38 &  $0$ & $0$ & $1$ & $2$ & $2$ & $2$ & $2$ \\
71 &  $0$ & $0$ & $0$ & $1$ & $1$ & $0$ & $0$ \\
72 &  $0$ & $0$ & $0$ & $1$ & $1$ & $2$ & $0$ \\
94 &  $0$ & $0$ & $0$ & $2$ & $1$ & $3$ & $0$ \\
140&  $0$ & $0$ & $1$ & $0$ & $1$ & $0$ & $0$ \\
\end{tabular}
\end{center}


\tablenum{1b}
\caption{This table lists the number of stars with colors satisfying
$1.52 < V-I < 2.26$ in the seven fields. There are seven histogram
bins each of 1 magnitude extent. The $I$ band magnitude centroids of
the bins are indicated at the top of each column. The starcounts in
the final column may be incomplete because of the upper limit on the $I$
band magnitude.}

\end{table*}

\eject

\begin{table*}
\begin{center}
\begin{tabular}{crrrrrc}
Model& $R_d$ & $F$ & $f$ & $\zthinz$ & $\zthickz$ & $\tau$ \\
\tableline
\null &\null&\null&\null&\null&\null&\null \\
A & $3.5$ & $10$ & $0.75$ & $350$ & $700$ & $3.8\times 10^{-8}$\\
\null &\null&\null&\null&\null&\null&\null \\
B & $4.5$ & $10$ & $0.90$ & $350$ & $1000$ & $4.4\times 10^{-8}$ \\
\null &\null&\null&\null&\null&\null&\null\\
C & $4.5$ & $5$ & $0.75$ & $350$ & $700$ & $2.3\times 10^{-8}$ \\
\null &\null&\null&\null&\null&\null&\null \\
\end{tabular}
\end{center}


\tablenum{2}
\caption{This table gives the parameters for the three
warped and flaring disk models examined in detail. They are the
scalelength $R_d$ and the scaleheights of the thin disk $\zthinz$ and
the thick disk $\zthickz$ at the solar radius. The flare parameter $F$
describes the factor by which the scaleheights increase at a
Galactocentric radii $25\,\kpc$. The final parameter $f$ describes the
relative proportion of the thin disk to the thick disk. For each
model, the microlensing optical depth $\tau$ is given in the final
column.}

\end{table*}

\eject

\begin{table*}
\begin{center}
\begin{tabular}{crrrrrrrrrr}
Field No & Model & $18$ & $19$ & $20$ & $21$ & $22$ & $23$ & $24$ \\
\tableline
4  & A & $0$ & $4$ & $10$ & $18$ & $27$ & $30$ & $1$ \\ 
36 & A & $0$ & $0$ & $3$ & $4$ & $2$ & $8$ & $17$ \\ 
38 & A & $0$ & $0$ & $3$ & $4$ & $2$ & $8$ & $17$ \\ 
71 & A & $0$ & $0$ & $9$ & $17$ & $28$ & $34$ & $0$ \\ 
72 & A & $0$ & $0$ & $5$ & $17$ & $28$ & $34$ & $5$ \\ 
94 & A & $0$ & $0$ & $3$ & $4$ & $2$ & $4$ & $6$ \\ 
140& A & $0$ & $0$ & $2$ & $3$ & $8$ & $29$ & $21$ \\ 
\null&\null&\null&\null&\null&\null&\null&\null&\null \\
4  & B & $0$ & $4$ & $10$ & $18$ & $31$ & $42$ & $2$ \\ 
36 & B & $0$ & $0$ & $2$ & $5$ & $4$ & $16$ & $31$ \\ 
38 & B & $0$ & $0$ & $2$ & $5$ & $4$ & $16$ & $31$ \\ 
71 & B & $0$ & $0$ & $8$ & $16$ & $31$ & $46$ & $0$ \\ 
72 & B & $0$ & $0$ & $5$ & $16$ & $31$ & $46$ & $8$ \\ 
94 & B & $0$ & $0$ & $2$ & $5$ & $4$ & $11$ & $14$ \\ 
140& B & $0$ & $0$ & $2$ & $4$ & $13$ & $43$ & $35$ \\ 
\null&\null&\null&\null&\null&\null&\null&\null&\null \\
4  & C & $0$ & $3$ & $5$ & $9$ & $13$ & $14$ & $1$ \\ 
36 & C & $0$ & $0$ & $2$ & $3$ & $2$ & $1$ & $2$ \\ 
38 & C & $0$ & $0$ & $2$ & $3$ & $2$ & $1$ & $2$ \\ 
71 & C & $0$ & $0$ & $5$ & $8$ & $13$ & $15$ & $0$ \\ 
72 & C & $0$ & $0$ & $3$ & $8$ & $13$ & $15$ & $2$ \\ 
94 & C & $0$ & $0$ & $3$ & $4$ & $2$ & $1$ & $0$ \\ 
140& C & $0$ & $0$ & $2$ & $3$ & $3$ & $6$ & $4$ \\ 
\end{tabular}
\end{center}


\tablenum{3}
\caption{This table lists the expected number of stars with colors satisfying
$1.52 < V-I < 2.26$ in the seven fields for each of the three
warped and flaring disk models.}

\end{table*}

\eject

\begin{table*}
\begin{center}
\begin{tabular}{crrrrrrrrrr}
Field No & $18$ & $19$ & $20$ & $21$ & $22$ & $23$ & $24$ \\
\tableline
4  & $0$ & $3$ & $6$ & $9$ & $8$  & $0$ & $0$ \\
36 & $0$ & $0$ & $3$ & $4$ & $2$  & $0$ & $0$ \\
38 & $0$ & $0$ & $3$ & $4$ & $2$  & $0$ & $0$ \\ 
71 & $0$ & $0$ & $6$ & $9$ & $10$ & $0$ & $0$ \\ 
72 & $0$ & $0$ & $4$ & $9$ & $10$ & $0$ & $0$ \\ 
94 & $0$ & $0$ & $3$ & $4$ & $2$  & $0$ & $0$ \\ 
140& $0$ & $0$ & $2$ & $3$ & $5$  & $3$ & $0$ \\ 
\end{tabular}
\end{center}


\tablenum{4}
\caption{This table lists the expected number of stars with colors satisfying
$1.52 < V-I < 2.26$ in the seven fields for model A assuming a metallicity gradient of 0.15 dex/kpc from the solar circle.}

\end{table*}

\eject

\end{document}